\documentclass[preprint]{elsarticle}
\usepackage{amssymb,amsbsy,amsmath,amsfonts,multirow}
\usepackage{yhmath}

\usepackage{soul}
\usepackage{slashed}
\usepackage{bm}
\usepackage{times}
\usepackage[normalem]{ulem}

\usepackage[dvipsnames,table,xcdraw]{xcolor}
\definecolor{lightyellow}{RGB}{255,250,205}

\usepackage{graphicx}
\usepackage{epstopdf}

\usepackage{tabularx}
\usepackage{longtable}

\usepackage{hyperref}
\hypersetup{
  colorlinks=true,        
  linkcolor=blue,         
  citecolor=cyan,         
  urlcolor=red,           
}
\makeatletter
\def\ps@pprintTitle{%
  \let\@oddhead\@empty
  \let\@evenhead\@empty
  \def\@oddfoot{\reset@font\hfil\thepage\hfil}
  \let\@evenfoot\@oddfoot
}
\makeatother
\begin{document}

\title{A study of the $\phi N$  correlation function}
\author[inst1]{Luciano M. Abreu}
\ead{luciano.abreu@ufba.br}
\address[inst1]{Instituto de F\'isica, Universidade Federal da Bahia, 40210-340, Salvador, BA, Brazil.}
\author[inst2,inst4]{Philipp Gubler}
\ead{philipp.gubler1@gmail.com}
\address[inst2]{Advanced Science Research Center, Japan Atomic Energy Agency, Tokai, Ibaraki 319-1195, Japan.}
\author[inst3]{K.~P.~Khemchandani}
\ead{kanchan.khemchandani@unifesp.br}
\address[inst3]{Universidade Federal de S\~ao Paulo, C.P. 01302-907, S\~ao Paulo, Brazil.}
\author[inst4]{A.~Mart\'inez~Torres}
\ead{amartine@if.usp.br}
 \address[inst4]{Universidade de Sao Paulo, Instituto de Fisica, C.P. 05389-970, S\~ao Paulo,  Brazil.}
\author[inst2,inst4,inst5]{Atsushi~Hosaka}
\ead{hosaka@rcnp.osaka-u.ac.jp}
 \address[inst5]{Research Center for Nuclear Physics (RCNP), Osaka University, Ibaraki 567-0047, Japan.}

\date{\today}

\begin{abstract}
The femtoscopic $\phi N$ correlation function is studied within a hadronic effective Lagrangian approach 
with coupled channels, based on hidden local symmetry. The results are compared 
with the data recently reported by the ALICE collaboration. 
We find that the correlation function has very different features for the $\phi N$ system in the spin 1/2 and  3/2 configurations, with the spin-averaged combination matching well with the experimental data. A  strong attraction, leading to the formation of a $\phi N$ bound state or a very prominent cusp is found in the spin 3/2 case. The correlation function for spin 1/2, on the other hand, 
is strongly impacted by the negative parity $N^*(1895)$ nucleon resonance.
\end{abstract}

\maketitle

\section{Introduction}
A first n\"aive guess on the nature of the $\phi$-$N$ interaction could be biased by the Okubo-Zweig-Iizuka (OZI) rule which would imply a weak interaction among the two hadrons due to their distinct quark contents. A more careful line of reasoning would, however, already raise questions on possible coupled channel effects. Indeed, 
the experimental study of correlations of $p -\phi \oplus \bar p -\phi $ pairs, measured in high-multiplicity $pp$ collisions at $\sqrt{s} = 13 $ TeV by the ALICE Collaboration~\cite{ALICE:2021cpv},
 indicates that the relevant interactions are far from those expected by the OZI suppression. 
A Lednický–Lyuboshits fit to the data led to a spin-averaged scattering length of $\left(0.85\pm 0.34\pm0.14\right)+i\left(0.16\pm0.1\pm0.09\right)$ fm, which was interpreted as an attractive interaction. 
Soon after the publication of Ref.~\cite{ALICE:2021cpv}, another piece of useful information on the $\phi N$ interaction was obtained from lattice QCD simulations based on the HAL QCD method~\cite{Lyu:2022imf}. Specifically, Ref.~\cite{Lyu:2022imf} reported values of the scattering length and the effective range for the $\phi N$ interaction in the spin-3/2 configuration as $a^{(3/2)}=-1.43(23)^{+36}_{-06}$ fm and $r^{(3/2)}_{\text{eff}} = 2.36(10)^{+02}_{-48}$fm, showing that the interaction is attractive. All these results seem to be at odds with the one inferred from the $\phi$ photoproduction data~\cite{Strakovsky:2020uqs}, which implies a weak $\phi N$ interaction as would be expected from the OZI rule. Using the findings of Refs.~\cite{ALICE:2021cpv,Lyu:2022imf}, a fit was made to the data on the correlation function in Ref.~\cite{Chizzali:2022pjd} by constraining the spin 3/2 interaction using the scattering length determined by  lattice QCD simulations~\cite{Lyu:2022imf}. As a result, evidence for a $\phi N$ bound state was found for the spin 1/2 channel. Yet another fit to the same data was recently presented in Ref.~\cite{Feijoo:2024bvn}, in which the correlation function was calculated using vector meson-baryon amplitudes obtained within a coupled channel approach based on hidden local symmetry (HLS). The interactions in this model proceed through a $t$-channel vector-meson exchange, which lead to  spin-independent correlation functions, and that differs from the findings of Ref.~\cite{Chizzali:2022pjd}.

In this paper, we present a different approach, where we do not make a fit, but simply use the previous results of Refs.~\cite{Khemchandani:2011et,Khemchandani:2011mf,Khemchandani:2013nma}. In Refs.~\cite{Khemchandani:2011et,Khemchandani:2011mf}, a formalism for meson-baryon interactions was developed by treating vectors and pseudoscalars on an equal footing up to all orders of the scattering equation. Besides, it was found that a contact interaction coming from the HLS Lagrangian and $u$-channel interactions gave contributions comparable to those coming from the $t$-channel diagrams. Consequently, the interactions were found to be both spin and isospin-dependent. Refs.~\cite{Khemchandani:2011et,Khemchandani:2011mf} were further extended in Ref.~\cite{Khemchandani:2013nma} by constraining the model parameters to reproduce some relevant experimental data on pseudoscalar-baryon final states. It is worth mentioning that the same model has been successfully applied to the study of several meson-baryon systems, predicting observables and reproducing experimental data when available~\cite{Khemchandani:2012ur,Khemchandani:2016ftn,Khemchandani:2020exc,Kim:2021wov,Khemchandani:2018amu}. Here, we will stay very close to the works of  Refs.~\cite{Khemchandani:2011et,Khemchandani:2011mf,Khemchandani:2013nma}, where it was shown that the amplitudes resulting from the coupled channel treatment exhibit a strong attraction near the $\phi N$ threshold in the spin 3/2 channel, while a weaker attraction is obtained in the spin 1/2 case (near the $\phi N$ threshold). As we will show, we find that such amplitudes lead to correlation functions which are different for the two spin cases, and are more in line with the findings of Ref.~\cite{Chizzali:2022pjd}.

The nature and strength of the $\phi$-$N$ interaction is of interest also in the context of studying in-medium behavior of the $\phi$ meson 
in nuclear matter \cite{Hatsuda:1991ez,Aoki:2023qgl,Gubler:2014pta} and the possibility of the formation of a $\phi$-nucleus bound state \cite{Gao:2000az,Cobos-Martinez:2017woo,Sun:2022cxf,Kuros:2024dhc,Filikhin:2024avj,Filikhin:2024xkb}, for which the state discussed 
in Ref.~\cite{Chizzali:2022pjd} would be the most elementary type. While the KEK E325 Collaboration reported a negative mass shift of $-3.4^{+0.6}_{-0.7}$\% for the $\phi$ meson in nuclear matter \cite{KEK-PS-E325:2005wbm}, the values of the above recently measured 
scattering lengths around 1 fm translate, within the linear density approximation, to mass shifts of the order of 10\% \cite{Gubler:2024ovg} 
(see Ref.~\cite{Paryev:2022zkt} for a discussion of some of the effects that go beyond linear density).
The J-PARC E16 \cite{Aoki:2023qgl} and E88 \cite{KEK:Sako2022} experiments will hopefully shed new light on this issue by providing 
updated measurements of the $\phi$ meson mass shift in nuclear matter. This paper will tackle the problem from the opposite side, 
by giving an improved interpretation of the ALICE $\phi N$ correlation function data based on a phenomenologically 
successful coupled channel approach, as described in the previous paragraph. 

This paper is organized as follows. We start by giving a brief outline of the model we use to determine meson-baryon amplitudes and the formalism to apply those amplitudes for computing the $\phi N$ correlation function. Finally, we present the main results of this work, the $\phi N$ correlation function and its comparison with the available experimental data. We furthermore provide the decomposition of the spin-averaged correlation function into different spins and channels and discuss the corresponding results.
The paper is concluded with a summary.
 
 \section{Formalism}
 \subsection{Meson-baryon scattering amplitudes}
 In this section, we first briefly discuss the calculation of the $\phi$-$N$ amplitudes which result from solving the Bethe-Salpeter equation. For more details we refer the reader to Refs.~\cite{Khemchandani:2011et,Khemchandani:2011mf,Khemchandani:2013nma}. The amplitudes were obtained in these former works by considering the following meson-baryon systems to build the coupled channel space in spin 1/2: $\pi N$, $\eta N$, $K\Lambda$, $K\Sigma$, $\rho N$, $\omega N$, $\phi N$, $K^*\Lambda$, and $K^*\Sigma$. It is important to mention that all the amplitudes are projected on the $s$-wave. In such a framework, only vector-baryon systems can couple to total spin 3/2 and, thus, the number of coupled channels is reduced in this case.
 The lowest-order vector-baryon interactions in our formalism are determined, through the Lagrangian~\cite{Khemchandani:2011et,Khemchandani:2013nma}
 \begin{eqnarray}\nonumber
&\mathcal{L}_{\textrm VB}& = -g \Biggl\{ \langle \bar{B} \gamma_\mu \left[ V_8^\mu, B \right] \rangle\!+\! \langle \bar{B} \gamma_\mu B \rangle  \langle  V_8^\mu \rangle  
\Biggr. + \frac{1}{4 M}\! \biggl( F \langle \bar{B} \sigma_{\mu\nu} \left[ V_8^{\mu\nu}, B \right] \rangle \! \biggr.\\
&&+\biggl.\! D \langle \bar{B} \sigma_{\mu\nu} \left\{V_8^{\mu\nu}, B \right\} \rangle\biggr)
 +  \Biggl.  \langle \bar{B} \gamma_\mu B \rangle  \langle  V_0^\mu \rangle  
+ \frac{ C_0}{4 M}  \langle \bar{B} \sigma_{\mu\nu}  V_0^{\mu\nu} B  \rangle  \Biggr\}, \label{vbb}
\end{eqnarray}
where the subscripts $8$ and $0$ represent the flavor octet and the singlet part of the wave function of the vector mesons. Such a separation is relevant for $\omega$ and $\phi$ which are considered as ideally mixed states of the octet and singlet components. The tensor field, $V^{\mu\nu}$, is written as 
\begin{equation}
V^{\mu\nu} = \partial^{\mu} V^\nu - \partial^{\nu} V^\mu + ig \left[V^\mu, V^\nu \right], \label{tensor}
\end{equation}
 with $V^\mu$ being the SU(3) matrix for the (physical) vector mesons

\begin{figure}[ht!]
    \centering
    \includegraphics[width=0.5\textwidth]{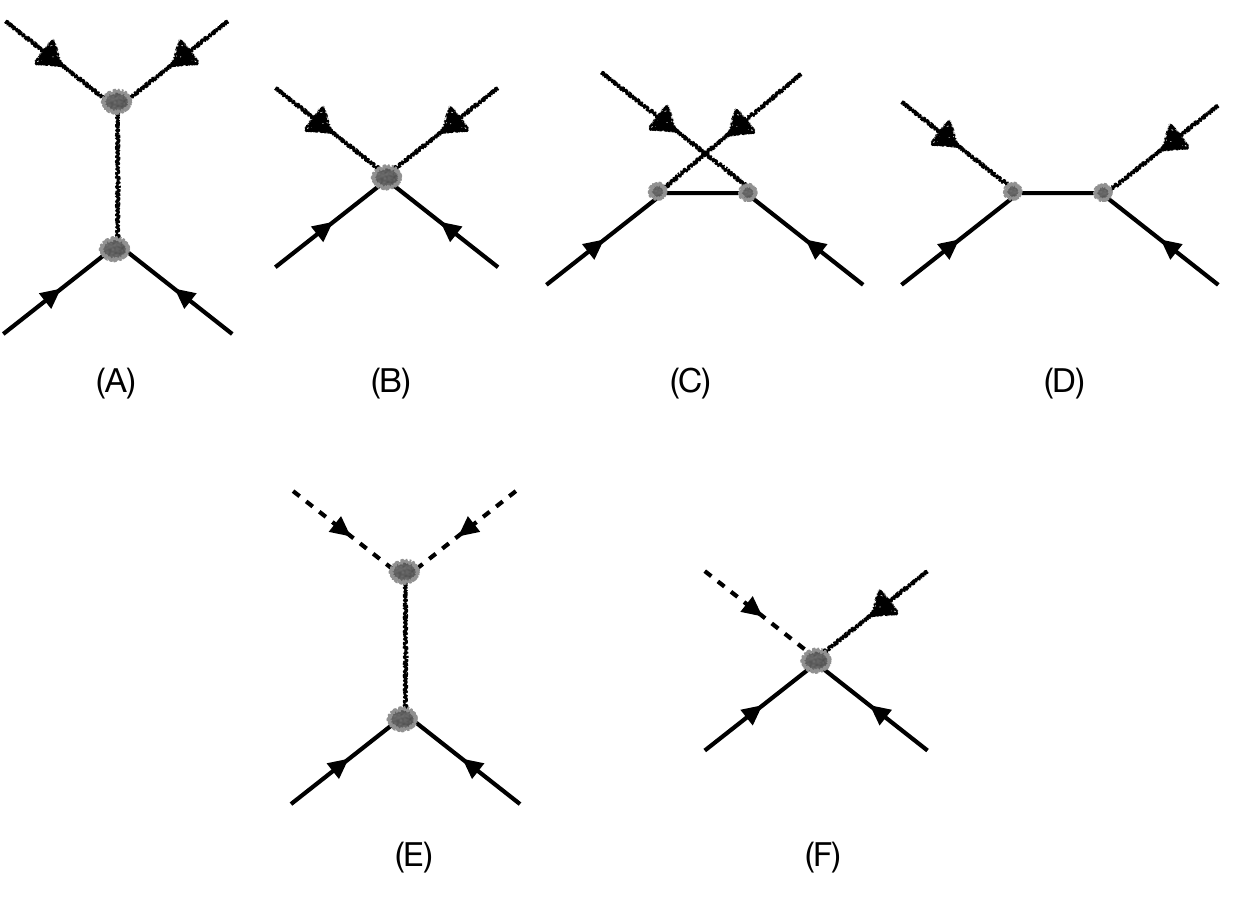}
    \caption{The upper row of diagrams correspond to the lowest order contributions to the vector-baryon interactions. The lower row shows the lowest order pseudoscalar-baryon amplitude and the one for the transition between the two types of systems. We represent baryons as solid lines, vector mesons as thick smeared lines and pseudoscalar mesons as dashed lines. }
    \label{diagram}
\end{figure}

 Four types of diagrams are determined from Eq.~(\ref{vbb}), which are shown in Fig.~\ref{diagram}(A)-(D). The spin part of the amplitudes corresponding to these four diagrams turns out, in the order the diagrams are shown,  to be (A)~$\vec\epsilon_i\cdot\vec\epsilon_f$, (B)~$\vec\sigma\cdot\vec\epsilon_f\times\vec\epsilon_i$, (C)~$\vec\epsilon_i\cdot\vec\sigma~\vec\epsilon_f\cdot\vec\sigma$ and (D)~$\vec\epsilon_f\cdot\vec\sigma~\vec\epsilon_i\cdot\vec\sigma$, where $\vec\epsilon_{i/f}$ are the polarization vectors 
 of the participating vector meson, while $\vec\sigma$ operates in the bayonic spin space. 

The pseudoscalar-baryon interactions in Refs.~\cite{Khemchandani:2011et,Khemchandani:2011mf,Khemchandani:2013nma} were determined from the lowest-order chiral Lagrangian (represented by diagram (E) in Fig.~\ref{diagram}). The transitions among the two types of meson-baryon systems were determined by introducing vector-mesons as gauge bosons in the nonlinear sigma model Lagrangian. We can interpret such amplitudes as an extension of the Kroll-Ruddermann term 
for the pion-photoproduction where a photon is replaced by a vector meson, relying on the vector meson dominance. 

We next discuss how these amplitudes are used as an input in the calculation of the correlation function.

\subsection{Correlation Function}\label{sec-cf}


The two-particle correlation function (CF) is formally defined as the ratio of the probability of measuring the two-particle state and the product of the probabilities of measuring each individual particle. After the use of some approximations (see a detailed discussion, for example, in Ref.~\cite{Lisa:2005dd}), it may be written in terms of the so-called Koonin–Pratt formula~\cite{Lisa:2005dd,Koonin:1977fh,Pratt:1986cc,Lednicky:1981su,Lednicky:1998}
\begin{eqnarray}
C(k) & = & \int d^3 r S_{12}(\vec{r}) \vert \Psi (\vec{k} ; \vec{r}) \vert ^2, 
\label{cf1}
\end{eqnarray}
where $ \vec{k} $ is the relative momentum in the center of mass (CM) frame of the pair; $\vec{r}$ is the relative distance between the two particles; $\Psi (\vec{k} ; \vec{r})$ is the relative two-particle wave function; and $ S_{12}(\vec{r}) $ is the source function.

For our practical purposes here, to accommodate the multi-channel amplitudes presented in the previous section, we  use a generalized version of the Koonin–Pratt formula~(\ref{cf1}) developed in preceding works~\cite{Vidana:2023olz,Feijoo:2023sfe,Albaladejo:2023pzq,Khemchandani:2023xup}. Accordingly, the coupled-channel CF for a specific channel $i$ is given by
\begin{eqnarray}
&& C_i(k_i) = 1 + 4 \pi \theta (q_{max} - k_i) \nonumber \\ 
&&\times \int_{0}^{\infty} d r r^2 S_{12}(\vec{r}) \left( \sum_j w_j \vert j_0(k_ir) \delta_{ji} + T_{ji}(\sqrt{s}) \widetilde{G}_j(r; s) \vert^2 - j_0^2(k_ir) \right) , 
\label{cf2}
\end{eqnarray}
where $ w_j $ is the weight of the observed channel $ j $; $ j_{\nu}(k_ir) $ is the spherical Bessel function; $E = \sqrt{s}$ is the CM energy; $k_i = \lambda^{1/2} (s,m_{1i}^2,m_{2i}^2)/(2\sqrt{s})$ is the relative momentum of the channel $i$, with $\lambda$ being the K\"allen function and $m_{1i}, m_{2i}$ the masses of the mesons in channel $i$; $ T_{ji} $ are the elements of the scattering matrix for the meson–baryon interactions; and the $\widetilde{G}_j(r; s)$ function is defined as 
\begin{eqnarray}
\widetilde{G}_j(r; s) & = & \int\limits_{\vert \vec{q} \vert < q_{max} } \frac{d^3 q}{(2\pi)^3} \frac{\omega_1^{(j)} + \omega_2^{(j)} }{2 \omega_1^{(j)} \omega_2^{(j)} } \frac{j_0(qr)}{s - \left( \omega_1^{(j)} + \omega_2^{(j)} \right)^2 +i \varepsilon} ,   
\label{gtilde}
\end{eqnarray}
with $ \omega_a^{(j)} \equiv \omega_a^{(j)}(q) = \sqrt{q^2 + m_a^2}$ being the energy of the particle $a$ in the channel $j$, and $q_{max}$ being a sharp cutoff momentum introduced to regularize the $r \to 0$ behavior.  The value of $q_{max}$ is chosen to be $q_{max} = 700 \  \mathrm{MeV}$. 

As discussed in the previous section, the $\phi N$ system is described by the spin states $1/2$ and $3/2$. Therefore, each spin contribution should be weighted by the spin degeneracy, and the total $\phi N$ CF is given by
\begin{eqnarray}
C_{\phi N}(k) & = & \frac{1}{3} C_{\phi N}^{\left(\frac{1}{2}\right)}(k) +  \frac{2}{3} C_{\phi N}^{\left(\frac{3}{2}\right)}(k), 
\label{cf3}
\end{eqnarray}
where $C_{\phi N}^{(S)}(k)$ is estimated from Eq.~(\ref{cf2}) making use of the $T$-matrices determined in the corresponding spin $s$ configuration.

Another important ingredient to be considered is the $r$-dependence of the source function. We employ the same parametrization adopted by the ALICE Collaboration  in Ref.~\cite{ALICE:2021cpv}. 
It is based on a static Gaussian profile normalized to unity, i.e.,
\begin{eqnarray}
S_{12}(\vec{r})  & = & \frac{1}{\left(  4 \pi \right)^{\frac{3}{2}} R^3} \exp{\left(  -\frac{r^2}{4 R^2 }\right)}. 
\label{sourcef1}
\end{eqnarray}
The source size parameter $R$ in the present work is fixed at the central value used in Ref.~\cite{ALICE:2021cpv}: $R = 1.08$ fm. A discussion on using other types of source functions can be found in Refs.~\cite{Kuroki_2024,Lisa:2005dd}.

The last elements to be remarked are the weights $w_j$'s in Eq.~(\ref{cf2}). In this regard, we benefit from the procedure reported in Ref.~\cite{Feijoo:2024bvn}, where each contribution associated to the $j$-th transition has been weighted considering the data-driven method used in Ref.~\cite{ALICE:2022yyh}. Accordingly, a given $w_j$ is related to the multiplicity of the pairs yielded from primary particles generated in the collision. In turn, this amount of pairs has been obtained in Ref.~\cite{ALICE:2022yyh} using the Thermal-Fist (TF) package~\cite{Vovchenko:2019kes,Vovchenko:2019pjl}, which is based on the statistical thermal model. The final number of pairs in the channel $j$ is then the product between the primary yields of the
particles in the considered pair, but taking into account pairs with relative momentum $k < 200 $ MeV in the Monte Carlo simulations of the kinematic distributions. Thus, to make a reasonable comparison, the weights employed here in the calculations of the $\phi N$ CF for the channel space in spin $ 3/2$ are the same as those in Ref.~\cite{Feijoo:2024bvn}. Concerning the remaining channels with spin $1/2$, we have used the findings of Refs.~\cite{Vovchenko:2019kes,Vovchenko:2019pjl}. The $w_j$'s are shown in Table~\ref{table-weights}, 
normalized with respect to the total production of $ \phi N $ pairs. 
\begin{table}[ht!]
\caption{Weights $w_j$'s for the corresponding channels employed in the calculation of the coupled-channel $\phi N$ CF shown in Eq.~(\ref{cf3}).  The $w_j$'s are
normalized with respect to the total production of $ \phi N $ pairs.}
\centering 
\begin{tabular}{ccc}
\hline
$j$-th channel & $w_j^{\left(\frac{1}{2}\right)}$ & $w_j^{\left(\frac{3}{2}\right)}$  \\
\hline 
$\pi N$                 & $ 71  $ & $ - $ \\
$ \eta N $              & $ 1   $ & $ - $ \\
$ K \Lambda $           & $ 5   $ & $ - $ \\
$ K \Sigma $            & $ 5   $ & $ - $ \\
$ \rho N $              & $ 6.24$ & $ 6.24$ \\
$ \omega N $            & $ 5.77$ & $ 5.77$ \\
$ \phi N $              & $ 1   $ & $ 1   $ \\
$ K^{\ast} \Lambda $    & $ 0.65$ & $ 0.65$ \\
$ K^{\ast} \Sigma $     & $ 0.42$ & $ 0.42$ \\
\hline 
\label{table-weights}
\end{tabular}
\end{table}

\section{Results}\label{subsec-cf-results}
\begin{figure}[ht!]
    \centering
\includegraphics[width=0.5\textwidth]{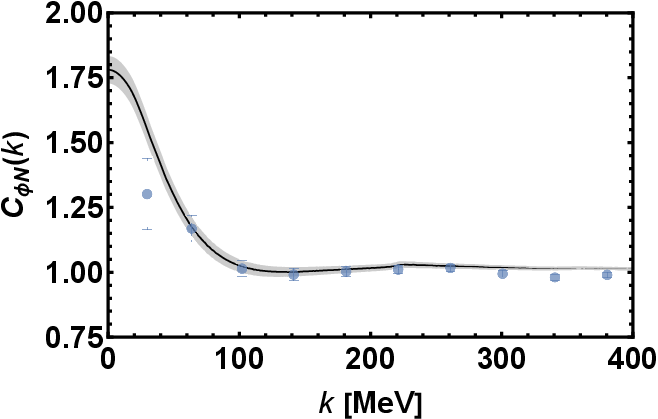}
\caption{Total coupled-channel $\phi N$ CF defined in Eq.~(\ref{cf3}) as a function of the CM relative momentum $k$.  The experimental points have been taken from Ref.~\cite{ALICE:2021cpv}. As can be noticed, the results are shown in the form of a band which is associated with the uncertainty in the values of the weights given in Table.~\ref{table-weights}. We consider the uncertainty to be of the order of $\pm 10\%$.} 
    \label{fig:compare_s3h_amps}
\end{figure}

\begin{figure}[ht!]
    \centering
\includegraphics[width=0.5\textwidth]{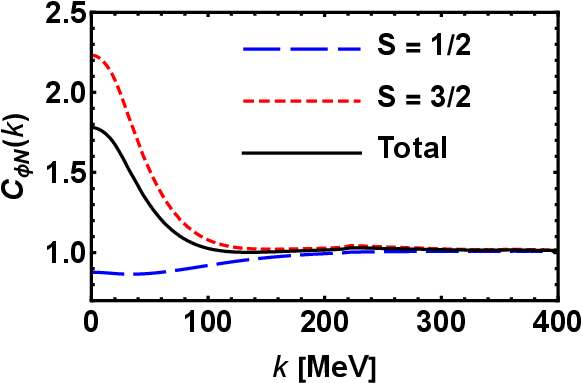}
\caption{The decomposition of the result shown in Fig.~\ref{fig:compare_s3h_amps} into its $S=1/2$ and $S=3/2$ components. }
    \label{fig:cf1}
\end{figure}

We begin by showing in Fig.~\ref{fig:compare_s3h_amps} the spin averaged correlation function obtained through Eq.~(\ref{cf3}), as a function of the CM relative momentum $k$. 
The calculation was done, for each spin case, using Eq.~(\ref{cf2})
with the weights listed in Table~\ref{table-weights}. 
For gaining a better understanding of this result, it is instructive to decompose it into its spin components. 
This is done in Fig.~\ref{fig:cf1}, where we plot the $\phi N$ CF determined for the two spin channels 1/2 and 3/2. 
Clearly, the spin 3/2 amplitudes give a positive and much larger contribution to the full (spin-averaged) correlation function, 
while the corresponding spin 1/2 curve starts below unity and shows a much smaller momentum dependence. 
We note that this behavior is in qualitative agreement with what was found in Ref.~\cite{Chizzali:2022pjd}. 
In our approach, this agreement can only be achieved once spin-dependent interactions 
and all relevant coupled channels are taken into account (as will be shown below). 
Furthermore, we do find a strong attraction in the spin 3/2 case, in accordance with Ref.~\cite{Lyu:2022imf}. 

\begin{figure}[h!]
    \centering
\includegraphics[width=0.5\textwidth]{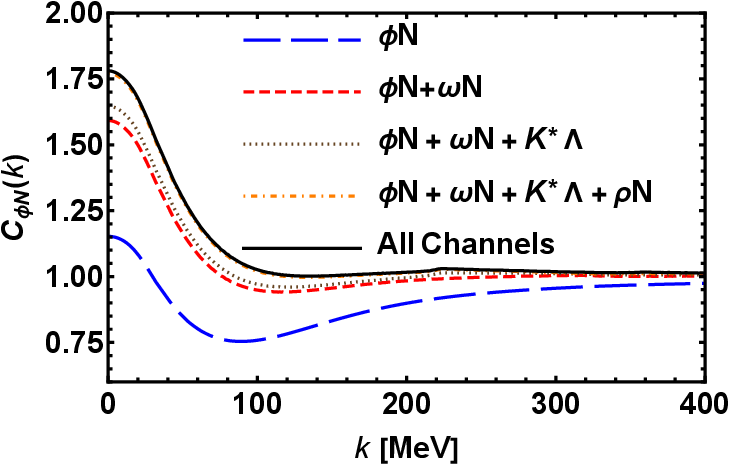}
\caption{Contribution  of different channels to the $\phi N$ CF  defined in Eq.~(\ref{cf3}). }
    \label{fig:cf2}
\end{figure}

\begin{figure}[ht!]
    \centering
\includegraphics[width=0.48\textwidth]{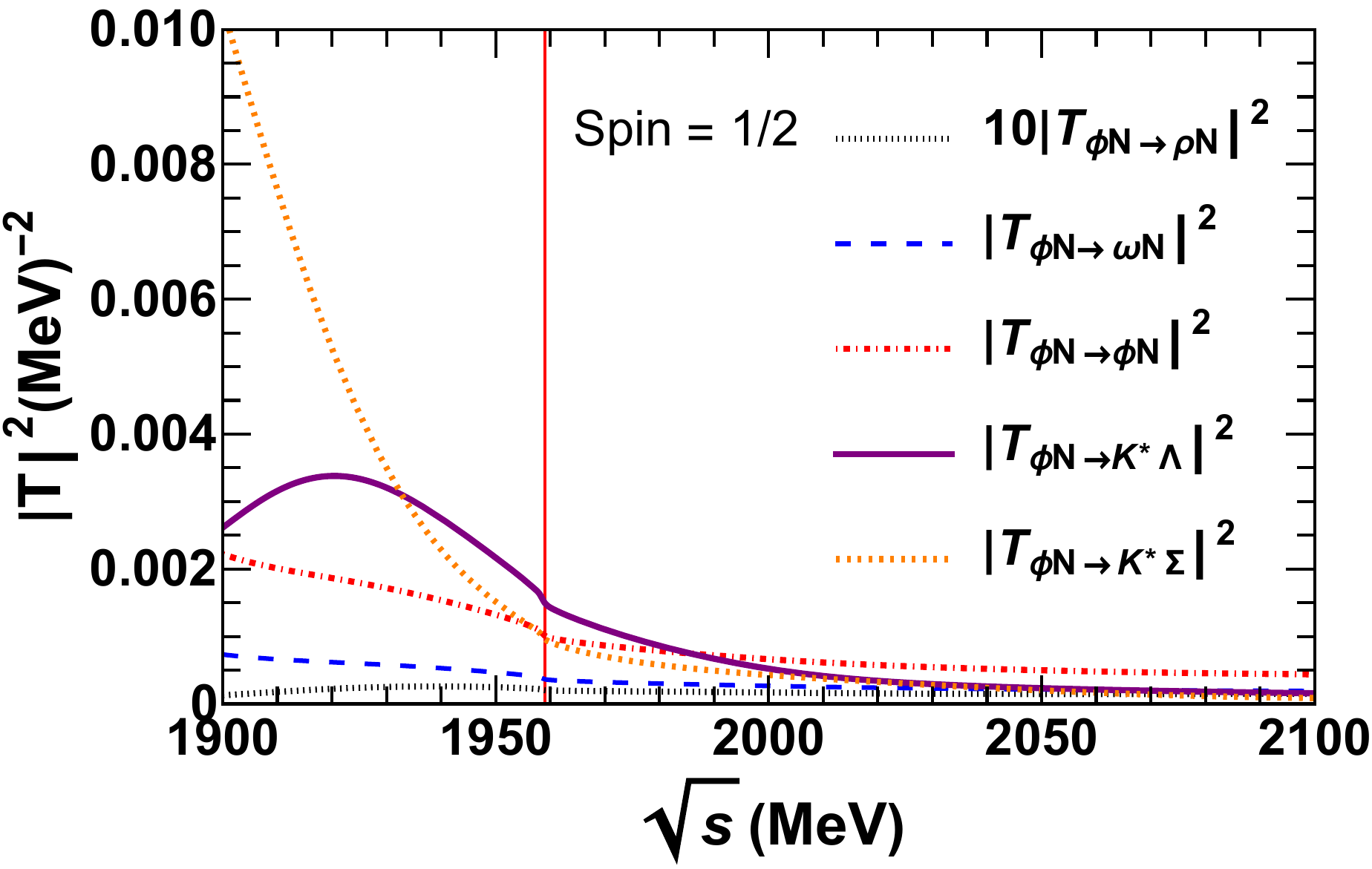} 
\includegraphics[width=0.48\textwidth]{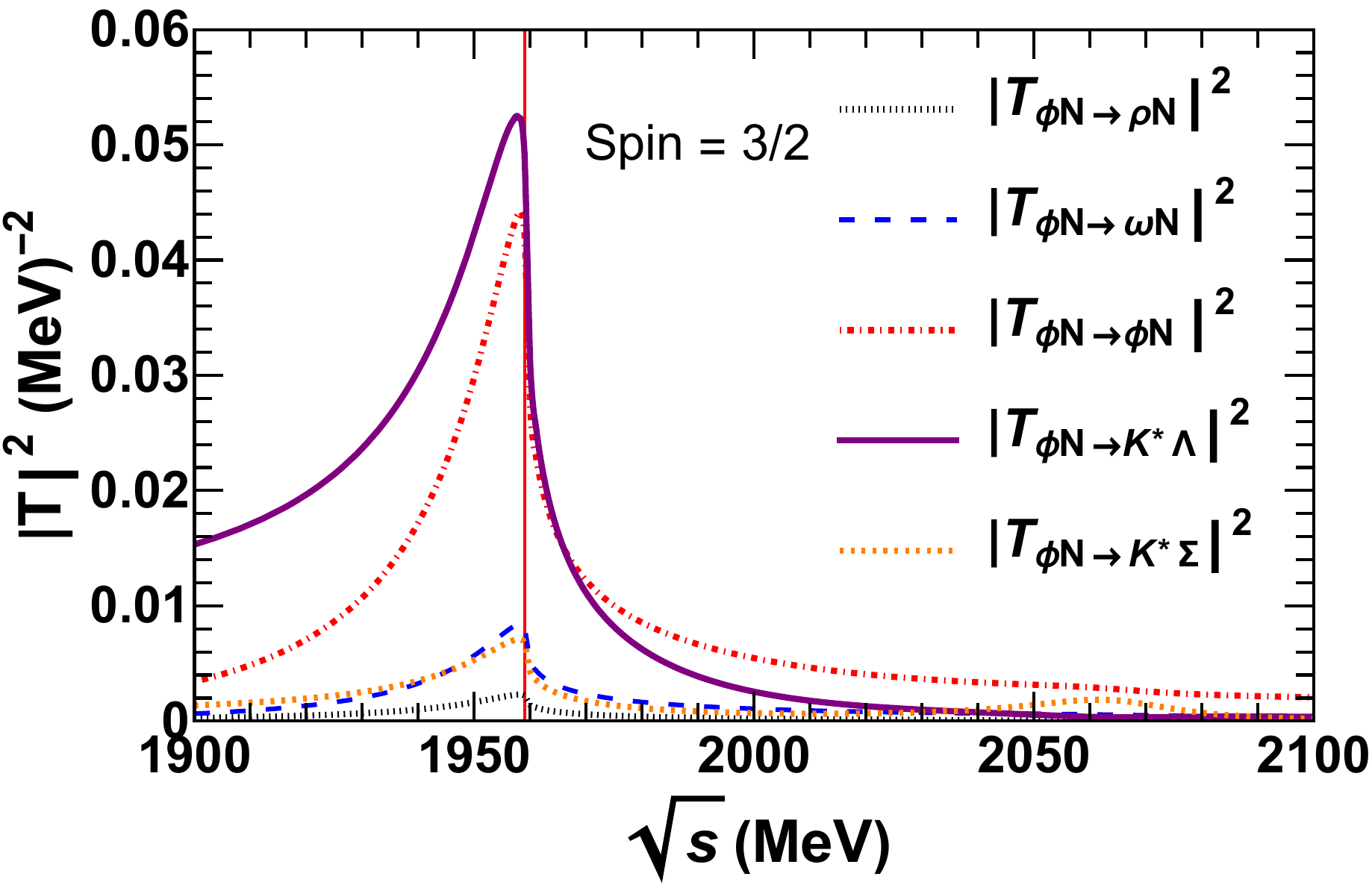}
\caption{This figure shows a comparison of the different amplitudes which are contributing to the CF in the spin 1/2 (left panel) and spin 3/2 (right panel) case. It can be seen that the dominant contribution comes from the $\phi N$ and $K^*\Lambda$ channels.} 
\label{fig:channel_amps}
\end{figure}

To understand this result further, we need to consider the spin 3/2 and 1/2 $\phi N$ amplitudes in detail. 
First, we note that solving the Bethe-Salpeter equation in the approach followed in 
Refs.~\cite{Khemchandani:2011et,Khemchandani:2011mf,Khemchandani:2013nma} involves regularizing a divergent loop function. This was done by following the dimensional regularizaton scheme,  
 using a natural size~\cite{Hyodo:2008xr} for the subtraction constants, $a_i=-2$, for each channel, $i$, in Ref.~\cite{Khemchandani:2011et}. The study on the strangeness zero meson-baryon systems was further extended in Ref.~\cite{Khemchandani:2013nma} by including pseudoscalar-baryon systems in the formalism as coupled channels in the spin 1/2 configuration. In Ref.~\cite{Khemchandani:2013nma}, the regularization   parameters of the model were constrained to fit the experimental data on processes involving pseudoscalar-baryon systems. 
Hence, the subtraction constants for the spin 1/2 case can be considered to be fixed and should not be varied. In the spin 3/2 case, though, it is possible to vary slightly the parameters  to study the uncertainties present in this case. With this in mind,
we use two sets (A and B) of subtraction constants to calculate the spin 3/2 amplitudes. The respective values are listed in Table~\ref{parameters_s3h}. 
\begin{table}[h!]
    \centering
    \begin{tabular}{c|cc}\hline
    Channel ($i)$    & $a_i$ (Set~A)&  $a_i$ (Set~B)\\\hline
       $\rho N$  & $-2.0$& $-2.0$\\
       $\omega N N$  & $-2.0$& $-2.0$\\
       $\phi N$&$-1.7$&$-2.0$\\
       $K^*\Lambda$&$-2.1$&$-2.1$\\
       $K^*\Sigma$&$-2.0$&$-2.0$\\\hline
    \end{tabular}
    \caption{Values of the model parameters used to calculate spin-3/2 meson-baryon amplitudes.}
    \label{parameters_s3h}
\end{table}
As can be seen, both Set A and B are very close to the parameters used in Ref.~\cite{Khemchandani:2011et}. We have used one of them, Set A, to prepare  
Figs.~\ref{fig:compare_s3h_amps}, \ref{fig:cf1}, \ref{fig:cf2}, and \ref{fig:channel_amps}. As we will show, the two sets would lead to equivalent results. Set A corresponds to slightly reducing the attraction as compared to the work in Ref.~\cite{Khemchandani:2011et}, while Set B corresponds to slightly increasing the attraction.

\begin{figure}[ht!]
    \centering
\includegraphics[width=0.48\textwidth]{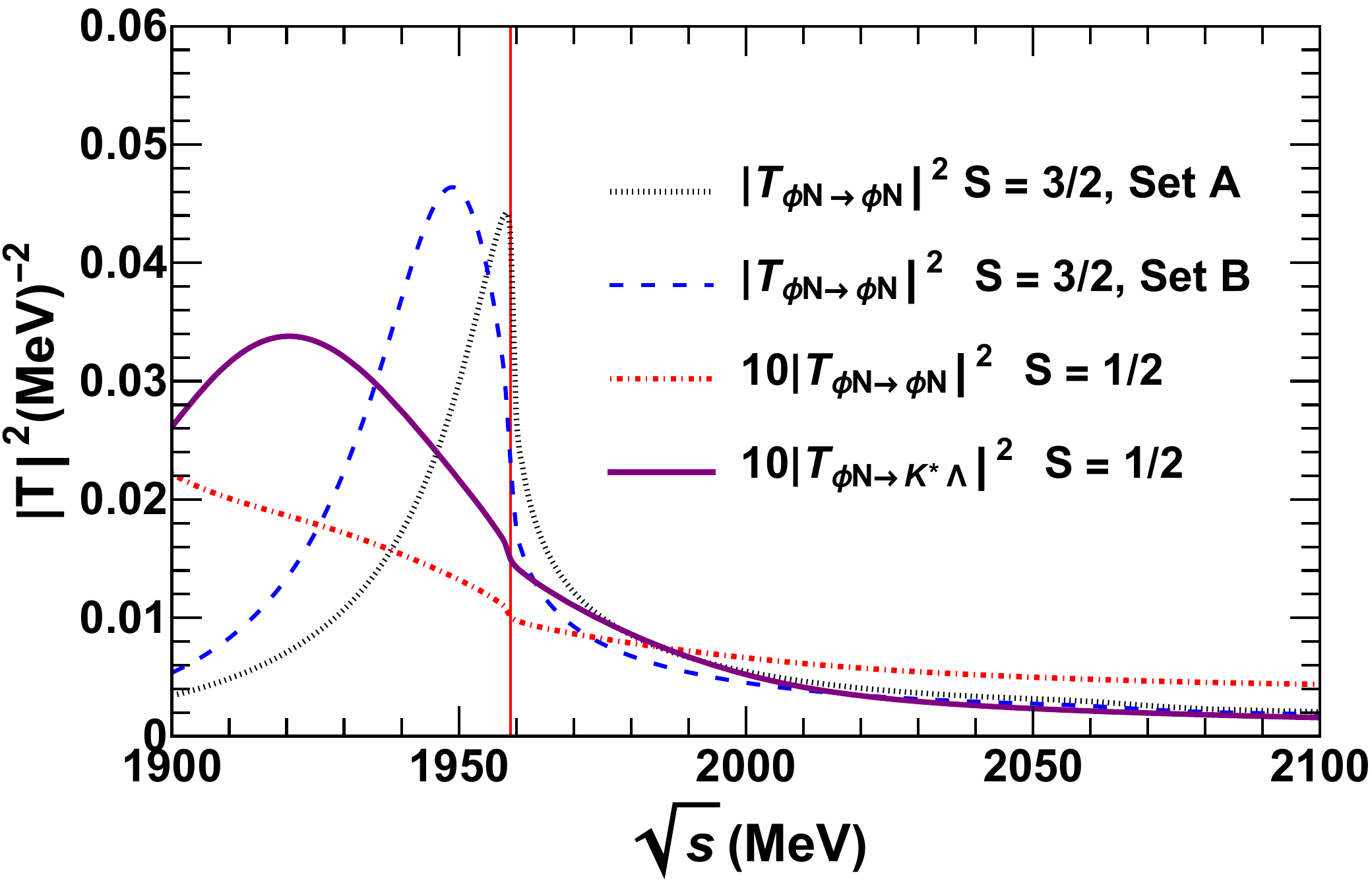}
\caption{A comparison of the amplitudes giving dominant contributions to the CF in the two spin configurations. We also show the effect of small changes made in the values of the parameters in the spin 3/2 case, which can turn the cusp near the $\phi N$ threshold (shown as the dotted line) to a ``bound" state (shown in the dashed line). The parameters are given in Table.~\ref{parameters_s3h}. Notice that a factor 10 is multiplied to the spin 1/2 amplitudes to facilitate a comparison with amplitudes in spin 3/2 case. } \label{fig:compare_spin_amp}
\end{figure}

Let us now look at the obtained $\phi N$ spin 1/2 and 3/2 amplitudes, shown together in Fig.~\ref{fig:compare_spin_amp}, for comparison. 
The spin 3/2 amplitudes show either a pronounced $\phi N$ cusp (set A) or a bound state (set B). It is important to mention here that a strong cusp was present in the work of Ref.~\cite{Khemchandani:2011et}.
Although the spin 3/2 amplitudes are different for the two parameter sets especially below the $\phi N$ threshold (vertical dashed line), 
the differences near (and above) the threshold are small, which implies that the correlation function is not sensitive to either 
choice of the parameters. 
Besides, it can be seen that the spin 1/2 amplitudes are much weaker around the $\phi N$ threshold, which explains 
the small momentum dependence in Fig.~\ref{fig:cf1}. For the spin 1/2 case, it is however 
worth mentioning the important role of the $N^*(1895)$ in this region. To illustrate this, we in the same figure 
depict the $\phi N \to K^*\Lambda$ amplitude in which the existence of the $N^*(1895)$ is clearly visible. We must remark here that the mass of $N^*(1895)$ is not necessarily 1895 MeV, which is just an average mass value. It is useful to remark that a similar behavior of the CF, related to a shallow bound state, $T_{cc}$ (and similar states) has been found in Refs.~\cite{Vidana:2023olz,Feijoo:2023sfe}.  It is also important to recall that the behavior of the CF can very much depend on the channel of observation, coupled to the same state and to the size of the source function~\cite{Fabbietti:2020bfg,Molina:2023jov,Sarti:2023wlg,Liu:2022nec,Liu:2023uly,Liu:2023wfo,Liu:2024nac}.

Comparing our results with those of Ref.~\cite{Chizzali:2022pjd}, where an indication of the existence of a spin 1/2 $\phi N$ bound state is found, 
we would like to mention that we are not sure if we can associate such a finding with the presence of $N^*(1895)$ in our amplitudes. It 
is a state which appears below the $\phi$-$N$ threshold, but cannot be interpreted as a zero-width $\phi N$ bound 
state in our formalism since it is coupled to several channels open for decay.

In this context, it is instructive to discuss the $\phi N$ scattering length values obtained in the different spin cases. 
Before giving the explicit values, we should mention that the normalization used in our work  to relate the scattering length with the $t$-matrix is
\begin{align}
    a_{\phi N}^{S}=-\frac{M_N}{4\pi\sqrt{s}}T_{\phi N}^{S},
\end{align}
where $M_N$ stands for the nucleon mass and the relation between the $T$-matrix and the amplitude $V$ obtained from the effective Lagrangian is $T=V+VGT$. The scattering length value in the spin 1/2 case was already given in  
Ref.~\cite{Khemchandani:2013nma}, but for the sake of completeness we here provide the values for both spin channels.
\begin{align}
    &a_{\phi N}^{S=1/2}=-0.22 + i0.00~\text{fm},\\
    &a_{\phi N}^{S=3/2, \text{set A}}=-0.30+i1.50~\text{fm},\\
   &a_{\phi N}^{S=3/2, \text{set B}}=-0.79+i0.83~\text{fm}.
\end{align}
Note that with the above conventions for $a$, $T$ and $V$, negative values of the real part of $a$ indicates either a repulsive interaction 
or formation of a shallow bound state below the threshold.
The scattering length values for the spin 3/2 case seem to be very sensitive to the values of the parameters given in Table~\ref{parameters_s3h}, which is a consequence of the presence of a strong cusp or a bound state very close to the $\phi N$ threshold. 
For the parameter set A, we get a spin averaged scattering length of $-0.22+i 1.50$~fm, while for set B we get $-0.72+i 0.83$~fm.

Next, we consider the contributions of the most significant channels to the obtained correlation function. 
The corresponding decomposition for the spin averaged case is shown in Fig.~\ref{fig:cf2}, 
It can be seen that the contributions from multiple channels have important contributions, the most important ones 
being $\phi N$ and $\omega N$.
The corresponding amplitudes are shown in Fig~\ref{fig:channel_amps}. 
One can see that  $\phi N$ and $K^*\Lambda$ couple strongly in the spin 3/2 case to produce a cusp (or almost a bound state-like) structure near the $\phi N$ threshold. In case of spin 1/2, we have the presence of $N^*(1895)$, which was shown to have a two-pole nature in Ref.~\cite{Khemchandani:2013nma}, with values $1801-i96$~MeV and $1912-i 54$~MeV. As different channels have different couplings to the two poles, different structures appear in the amplitudes on the real axis.

\section{Conclusions}
In this work, we have calculated the correlation functions for the different spin configurations of the $\phi N$ channel, 
taking into account spin-dependent vector-baryon interactions derived from effective theory approach based on hidden local 
symmetry and coupled channel effects. 
We find that the presence of a strong attraction near the $\phi N$ threshold in the spin 3/2 channel gives rise to a CF larger 
than 1 for small relative momentum. On the other hand, the spin 1/2 CF exhibits the opposite behavior, which we relate to the presence 
of the $N^*(1895)$ resonance in the scattering amplitudes. 
When combining the two spin channels to a spin averaged quantity, we find reasonable agreement with the data reported by 
the ALICE Collaboration \cite{ALICE:2021cpv}, which is achieved without any parameter fitting, but is rather a natural 
outcome of the model. 
We find that coupled channel effects are crucial in determining the CF. Furthermore, the weights $w_j$ for the different channels 
contributing to the CF are also a critical input necessary to reach a good agreement with the experimental data.

\section{Acknowledgements}

This study was financed in part by the Coordena\c c\~ao de Aperfei\c coamento de Pessoal de N\'ivel Superior – Brasil (CAPES) – Finance Code 001.
The partial support from other Brazilian agencies is also gratefully acknowledged. We thank CNPq (L.M.A.: Grant No. 309950/2020-1, 400215/2022-5, 200567/2022-5; K.P.K: Grants No. 407437/ 2023-1 and No. 306461/2023-4; A.M.T: Grant No. 304510/2023-8), FAPESP (K.P.K.: Grant Number 2022/08347-9; A. M. T.: Grant number 2023/01182-7); and CNPq/FAPERJ under the Project INCT-F\'{\i}sica Nuclear e Aplica\c c\~oes (Contract No. 464898/2014-5). 
P.G. is supported by the Grant-in-Aid for Scientific Research (A)  (JSPS KAKENHI Grant Number JP22H00122).

\bibliographystyle{elsarticle-num.bst}
\bibliography{phinBib}

\end{document}